\documentclass[prl,twocolumn,showpacs]{revtex4}
\usepackage{psfig}
\begin{document}
\def\fw{86mm}
\def\/{\over}
\def\<{\langle}
\def\>{\rangle}
\def\({\left(}
\def\){\right)}
\def\[{\left[}
\def\]{\right]}
\def\Re{\mbox{Re}}
\def\Im{\mbox{Im}}
\def\mod{\,\mbox{mod}\,}
\def\i{{\rm i}}
\def\e{{\rm e}}
\def\dxq{{\Delta_{x}^2}}
\def\delx{{\Delta_{x}}}
\def\dpq{{\Delta_{p}^2}}
\def\delp{{\Delta_{p}}}
\def\g{g}
\def\N{{\cal N}}
\def\I{{\cal I}}
\def\A{{\cal A}}
\title{Phase-space correlations of chaotic eigenstates} 
\author{Holger Schanz}
\email{holger@chaos.gwdg.de}
\affiliation{Max-Planck-Institut f\"ur Dynamik und Selbstorganisation und Institut
f{\"u}r Nichtlineare Dynamik der Universit{\"a}t G{\"o}ttingen, 
Bunsenstra{\ss}e 10, D-37073 G\"ottingen, Germany}
\date{\today}
\pacs{03.65.Sq, 05.45.Mt}
\begin{abstract}
  It is shown that the Husimi representations of chaotic eigenstates are
  strongly correlated along classical trajectories. These correlations extend
  across the whole system size and, unlike the corresponding eigenfunction
  correlations in configuration space, they persist in the semiclassical limit. A
  quantitative theory is developed on the basis of Gaussian wavepacket
  dynamics and random-matrix arguments. The role of symmetries is discussed
  for the example of time-reversal invariance.
\end{abstract}
\maketitle
Chaotic eigenfunctions and in particular their localization and correlation
statistics are a topic of continuing interest
\cite{Ber77,Mir00,Hel84,Pri95,FE96,HS98,Han98}. Applications include classical, mesoscopic 
and pure quantum systems such as
optical, mechanical and microwave resonators \cite{H+01b,S+03c,KKS95},
electron transmission and interaction in chaotic quantum dots
\cite{Alh00,B+99c}, and decay and fluctuations of heavy nuclei \cite{Z+96}.
One foundation of eigenfunction statistics is the random-wave model of Berry
\cite{Ber77} which is essentially equivalent to random-matrix theory (RMT)
\cite{Mir00}. Within RMT the eigenfunction components in an arbitrary basis
are uncorrelated Gaussians. Current research is frequently aiming at
deviations from RMT due to the specific dynamics. Prominent examples are
scarring by periodic orbits \cite{Hel84} or long-range correlations
\cite{HS98}.  As classical dynamics takes place in phase space,
representations of eigenstates via Husimi or Wigner functions seem appropriate,
and recently some of their statistical properties have attracted a lot of
attention \cite{Han98}.  Surprisingly this does not apply to dynamically induced
correlations although numerous studies of the corresponding {\em
  spatial} correlations demonstrate their relevance \cite{Pri95,HS98,FE96},
and although there are systems where a direct relation between phase-space
correlations and measurable quantities must be expected. For example, in
optical resonators \cite{H+01b} the power emitted at a certain point of the
boundary depends strongly on the angle of incidence of the wave (total
internal reflection). Also in quantum dots effects of eigenfunction
directionality have been measured using tilted leads \cite{B+99c}.

In this paper we analyze for the first time dynamical correlations between
points in phase space. Our results are surprising in view of the fact that 
{\em spatial} two-point correlators of eigenfunction amplitudes or
densities vanish in the semiclassical limit $\hbar\to 0$ for any $x\ne
x'$ \cite{fn1}.  In contrast we find strong and semiclassically persistent
correlations between phase-space points $\xi\ne \xi'$ ($\xi=(x,p)$) where the
distance between $x$ and $x'$ can be of the order of the system size. This is
no contradiction since $\<\psi(x)\psi(x')\>=0$ does not imply that $\psi(x)$
and $\psi(x')$ are statistically independent; it just means that there are no
{\em linear} correlations in configuration space. In other words, although the
existence of dynamically induced correlations cannot be a matter of the chosen
basis they turn out to be most relevant in phase space.

In our approach we make use of methods which proved successful in studies of
eigenfunction scarring. We will see that Gaussian wave packet dynamics
\cite{Hel91} can also be applied to study eigenfunction properties which are
not at all related to periodic orbits. As in nonlinear scarring \cite{Hel84}
we have to supplement short-time dynamics by RMT in order to get a complete
theory. However, the new context requires to do so in a technically different
way. 

\begin{figure}[tb]
 \centerline{
  \psfig{figure=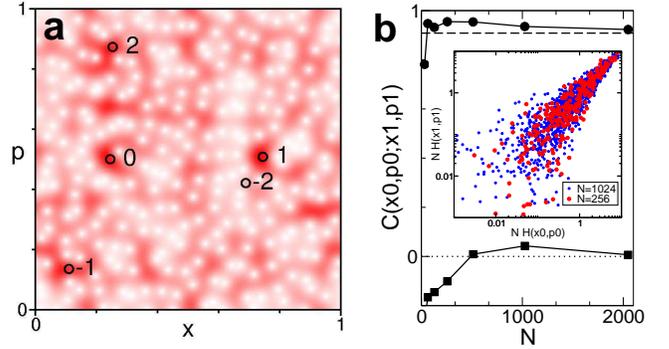,width=\fw}
 }
 \caption{\label{fig1} (a) An eigenstate of a chaotic torus map quantized 
   with $N=256$ is shown in Husimi representation. A classical trajectory
   $\xi_{t}=(x_{t},p_{t})$ of the same map is marked by circles with label
   $t$. Inset of (b): In a scatter plot the values of the Husimi density at
   $\xi_{0}$ and $\xi_{1}$ cluster near the diagonal which indicates strong
   correlation. (b) The correlation coefficient is shown with circles as
   function of $N$. The dashed line is the prediction of Eq.~(\ref{a1}).
   The squares show that spatial density correlations between the corresponding
   positions $x_{0}$ and $x_{1}$ vanish as $N\to\infty$.}
\end{figure}
We introduce our main result with a numerical example. Fig.~\ref{fig1}a
shows the Husimi representation of an eigenstate of a quantum kicked rotor,
\begin{equation}\label{propagator}
\hat U=\e^{-2\pi\i\,V(\hat x)/h}\,\e^{-2\pi\i\,T(\hat p)/h}\,.
\end{equation}
The unitary matrix $\hat U$ has dimension $N\equiv h^{-1}$ and
quantizes a
classical map on the torus ($x+1\equiv x$, $p+1\equiv p$),
\begin{equation}\label{map}
x_{t+1}=x_{t}+T'(p_{t})\qquad  p_{t+1}=p_{t}-V'(x_{t+1}).
\end{equation}
We choose $T(p)=p^2/2$ and a
potential for which this map is globally chaotic 
\cite{fn2}.
Therefore the eigenstates of $\hat U$ fill the entire phase space more or less
uniformly.  The fluctuations on this background are the object of our
interest. In Fig.~\ref{fig1}a we observe patches of high or low density
which are typically of the size of a Planck cell $h$. Besides this well-known
local correlation \cite{Han98}, the figure suggests the existence of
long-range correlations {\em along the trajectories of the underlying
  classical dynamics}. One such trajectory $\xi_{t}$ (not periodic!) is indicated by circles.
The point $\xi_{0}$ is close to a maximum of the density, and we observe
that also at the points $\xi_{t}$ the Husimi density has values above the
average, at least for $-1\le t\le 2$. A scatter plot of the densities at
$\xi_{0}$ and $\xi_{1}$ shows that this is no coincidence (inset of
Fig.~\ref{fig1}b). The corresponding correlation coefficient (circles in
Fig.~\ref{fig1}b) is as high as $0.94$ although the points
$\xi_{0}$ and $\xi_{1}$ are far apart and not related by any symmetry.  This
is in strong contrast to the absence of long-range correlations in position or
momentum representation in the semiclassical limit, which is
expected from previous studies \cite{Pri95,HS98,FE96} and confirmed in
Fig.~\ref{fig1}b for our model (squares).

The existence of dynamical correlations in phase space can be derived from
Gaussian wave packet dynamics \cite{Hel91}.  For the special case of the
quantum map Eq.~(\ref{propagator}) this standard approach is formulated as
follows.  The wave function of a general Gaussian state centered in phase
space at $\xi_{0}=(x_{0},p_{0})$ has the form
\begin{equation}\label{gauss}
\g(\xi_{0},\Delta \xi,\N,s;x)=\N
\e^{-\A_{\Delta\xi, s}
{(x-x_0)^2}+\i {p_0 (x-x_{0})/\hbar}}
\end{equation}
where $\A$, $\N$ are complex.  The prefactor $\N$ accounts for normalization and an overall
phase while  
\begin{equation}\label{A}
\A_{\Delta\xi,s}
=(2\delx)^{-2}
\({1+{s(2\i/\hbar)}\sqrt{\dxq\dpq-{\hbar^2\!/4}}}\)
\end{equation}
depends on the uncertainties of position and momentum,
$\Delta\xi=[\delx,\delp]$, and on a sign $s=\pm 1$. We assume that both,
$\Delta_{x}$ and $\Delta_{p}$, are much smaller than any relevant classical
scale 
\cite{fn3}.
In this case and
within the semiclassical approximation (stationary phase approximation for
$N\to\infty$) application of the propagator (\ref{propagator})
preserves the Gaussian form of the wave function,
\begin{equation}\label{propgauss}
\hat U\,|g(\xi_{0},\Delta\xi_{0},\N_{0},s_{0})\>
=
|g(\xi_{1},\Delta\xi_{1},\N_{1},s_{1})\>.
\end{equation}
In this equation $\xi_{1}$ is the classical iterate of $\xi_{0}$.
Position and momentum uncertainties transform as
\begin{eqnarray}
\label{transdx}
\dxq_{1}&=&
\dxq_{0}-2s_{0}T_{0}''\sqrt{\dxq_{0}\dpq_{0}-\hbar^2\!/4}+\dpq_{0}{T_{0}''}^2
\\
\label{transdp}
\dpq_{1}&=&\dpq_{0}
+2s_{0}'{V_{1}''}\sqrt{\dxq_{1}\dpq_{0}-{\hbar^2\!/4}}+\dxq_{1}{V_{1}''}^2
\end{eqnarray}
and also for $\N_{1}$ and $s_{1}$ semiclassical expressions are found. In a
chaotic system the growth of the uncertainties under repeated application of
Eqs.~(\ref{transdx}), (\ref{transdp}) will be dictated by the classical
Lyapunov exponent $\lambda$ after a short initial period because then
$\hbar^2\!/4\ll \dxq_{t}\dpq_{t}$ can be neglected and one is left with
equations that are essentially classical. Hence, for
$\dxq_{0},\dpq_{0}\sim\hbar$ the uncertainties grow to values $\sim 1$ after a
time of the order of the Ehrenfest time $t_{\rm E}\sim\lambda^{-1}\,\ln\hbar$ and
then the approximate Gaussian wave packet dynamics breaks down.

We can now project the semiclassical iterates
$|g_{t}\>\equiv|g(\xi_{t},\Delta\xi_{t},\N_{t},s_{t})\>$ of a
Gaussian $|g_{0}\>$ onto an eigenstate $|n\>$ of the propagator, $\hat
U\,|n\>=\exp(i\varepsilon_{n}/\hbar)\,|n\>$, and obtain
\begin{equation}\label{corr}
\<n|g_{t}\>=\exp(i\varepsilon_{n} t/\hbar)\,\<n|g_{0}\>
\end{equation}
for times up to the Ehrenfest time. Already from this simple equation
it is obvious that the projections of eigenstates into phase space
along a classical trajectory cannot be independent of each other and that
there must exist cross-correlations between eigenvalues and eigenstates. Note
that Eq.~(\ref{corr}) is a semiclassical identity which has to be satisfied
by every single eigenstate even without averaging. However, the
relevance of this equation is diminished by the fact that in general
$|g_{0}\>$ and $|g_{t}\>$ are Gaussian wave packets with completely different
widths and shapes. Therefore it is not immediately possible to extract
meaningful correlation functions relating, e.g., Husimi densities at different
points in phase space.  This problem will be addressed in the following.

\begin{figure}[tbp]
 \centerline{
  \psfig{figure=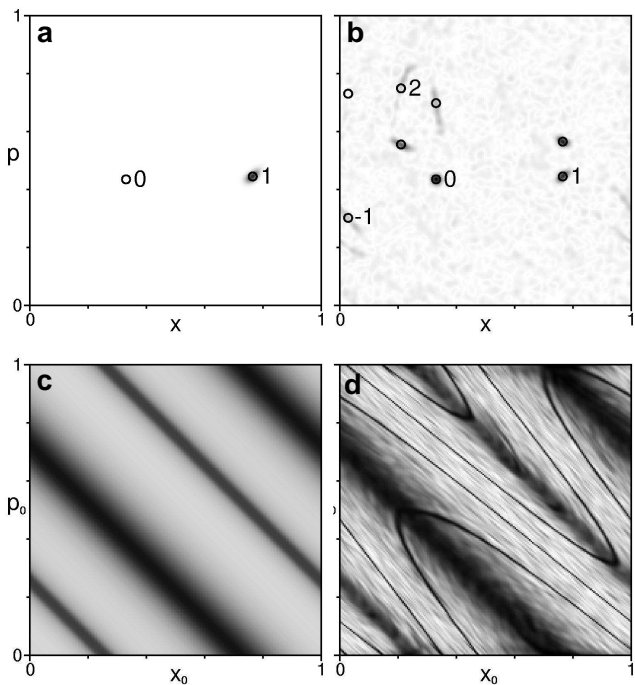,width=\fw}
 }
 \caption{\label{fig2} (a) For fixed $\xi_{0}$ the correlator of Husimi
   amplitudes $c_{1}(\xi_{0};\xi)$ is displayed ($N=1024$). The point
   $\xi_{0}$ and its classical iterate $\xi_{1}$ are marked by
   circles. (b) Same for the density correlator $C_{H}(\xi_{0};\xi)$.
   Labeled circles show the classical trajectory $\xi_{t}$. Unlabeled
   circles show the time-reversed trajectory $\tilde\xi_{t}$ obtained
   by the transformation (\ref{trpoint}). (c) $c_{1}(\xi_{0};\xi_{1})$
   is shown as a function of $\xi_{0}$. The magnitude of correlation is
   approximately constant along lines $x_{1}=x_{0}+p_{0}=$ const. (d) Same for
   $C_{H}(\xi_{0};\xi_{1})$. On top of a background depending on
   $x_{1}$ one observes smooth fluctuations and relatively sharp lines of
   excess correlation which correspond to Eq.~(\ref{line}).}
\end{figure}

For the sake of simplicity we will consider coherent states with 
equal position and momentum uncertainties,
\begin{eqnarray}\label{cohstate}
\alpha_{\xi_0}(x)
&=&(\hbar\pi)^{-1/4}\,\e^{
-{(x-x_0)^2}/2\hbar+\i {p_0} (x-x_{0})/\hbar}.
\end{eqnarray}
The Husimi amplitude of a state $|\psi\>$ at a point $\xi$ is
$h_{\psi}(\xi)=\<\alpha_{\xi}|\psi\>$ and the Husimi density is
$H_{\psi}(\xi)=|h_{\psi}(\xi)|^2$. We are mainly interested in the 
correlation coefficient (normalized covariance) for Husimi densities at two
different points $\xi$ and $\xi'$. It is
given by
\begin{equation}\label{denscorr}
C_{H}(\xi';\xi)={\<{\delta H_{n}(\xi)\,\delta H_{n}(\xi')}\>_{n}\over 
\sqrt{\<\delta^2 H_{n}(\xi)\>_{n}\<\delta^2 H_{n}(\xi')\>_{n}}}
\end{equation}
where $\<\cdot\>_{n}=N^{-1}\sum_{n=1}^{N}(\cdot)$ denotes the average over all
eigenstates $|n\>$ and $\delta H_{n}(\xi)=H_{n}(\xi)-N^{-1}$ is the deviation
of the Husimi density from its mean value $\<H_{n}(\xi)\>_{n}=N^{-1}$.  It
will be instructive to consider in parallel also a suitably defined
time-dependent correlator of Husimi amplitudes
\begin{equation}\label{ampcorr}
c_{t}(\xi';\xi)=
|N\<\e^{+\i\varepsilon_{n} t/\hbar}h_{n}^{*}(\xi')h_{n}(\xi)\>_{n}|^2
\end{equation}
which involves the eigenphase $\varepsilon_{n}$.  Fig.~\ref{fig2} illustrates
the main properties of the correlation functions (\ref{denscorr}),
(\ref{ampcorr}) which we shall explain below semiclassically. Fig.~\ref{fig2}a
shows for $t=1$ that $c_{t}(\xi_{0};\xi)$ as a function of $\xi$ is
concentrated at the classical iterate $\xi_{t}$. The magnitude of the
correlation between these two points depends on the classical trajectory. This
dependence is shown in Fig.~\ref{fig2}c by plotting
$c_{1}(\xi_{0};\xi_{1})$ as a function of $\xi_{0}$.  Figs.~\ref{fig2}b,d
contain the same information for the correlator $C_{H}$ whose structure is
more complicated. In Fig.~\ref{fig2}b we see that the Husimi density
along a classical trajectory $\xi_{t}$ is correlated over a short time, as
expected from Fig.~\ref{fig1}. In addition we observe also a strong
correlation to the time-reversed trajectory and some fluctuating background of
weakly positive or negative correlations covering the entire phase space. The
dependence of $C_{H}(\xi_{0};\xi_{1})$ on the position in phase space
(Fig.~\ref{fig2}d) resembles the characteristic pattern observed already for
the correlator of amplitudes in Fig.~\ref{fig2}c. However, it is blurred by
fluctuations and has in addition sharp lines of particularly high
correlation which we shall relate to the time-reversal symmetry of our model.

\begin{figure}[tb]
 \centerline{
  \psfig{figure=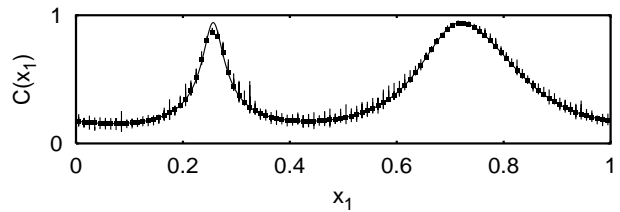,width=\fw}
 }
 \caption{\label{fig3} Amplitude and density correlators
   $c_{1}(\xi_{0};\xi_{1})$ (points) and $C_{H}(\xi_{0};\xi_{1})$ (error bars)
   are compared to the semiclassical prediction Eq.~(\ref{a1}) (solid line).
   An error bar shows mean value and standard deviation of those data from
   Fig.~\ref{fig2}d which are on the line $x_{1}=x_{0}+p_{0}$ but away
   from the exceptional curves given by Eq.~(\ref{line}).}
\end{figure}

It is straightforward to obtain a semiclassical approximation to
Eq.~(\ref{ampcorr}).  Using the completeness of eigenstates we find
$\sum_{n}\e^{+\i\varepsilon_{n} t/\hbar}h_{n}^{*}(\xi_{0})h_{n}(\xi)
=\<\alpha_{\xi}|\hat U^{t}|\alpha_{\xi_0}\>$. As the semiclassical limit is
approached, $\hat U^t|\alpha_{\xi_0}\>$ localizes at $\xi_{t}$ and the overlap
with $|\alpha_{\xi}\>$ vanishes for any other given point (Fig. 2a). Therefore all
relevant correlations are given by 
$c_{t}(\xi_{0};\xi_{t})=|a_{t}(\xi_{0})|^2$
with 
\begin{equation}\label{overlap}
a_{t}(\xi_{0})=\<\alpha_{\xi_t}|\hat U^{t}|\alpha_{\xi_0}\>.
\end{equation}
Finally, this overlap is approximated semiclassically with the help of
Eqs.~(\ref{transdx}), (\ref{transdp}). For example we find at $t=1$
\begin{equation}\label{a1}
|a_{1}(\xi_{0})|^2=2^{3/2}(9+[2V''(x_{0}+p_{0})-1]^2)^{-1/2}.
\end{equation}
Note that $c_{1}(\xi_{0};\xi_{1})=|a_{1}(\xi_{0})|^2$ depends only
on $x_{1}=x_{0}+p_{0}$ which explains the stripes in Fig.~\ref{fig2}c. In
Fig.~\ref{fig3} we check the accuracy of Eq.~(\ref{a1}) by comparing it to the
exact value for $N=1024$ (points vs solid line).

We continue with a semiclassical theory for the density correlator
Eq.~(\ref{denscorr}) which allows to understand the main features observed in
Figs.~\ref{fig2}b,d. For this purpose we rephrase Eq.~(\ref{overlap}) as
\begin{equation}\label{itercoh}
\hat U^{t}|\alpha_{0}\>=
a_{t}|\alpha_{t}\>
+r_{t}|\rho_{t}\>
\end{equation}
where $|\rho_{t}\>$ is a normalized state which is localized near
$\xi_{t}$ but orthogonal to the coherent state at this position,
$\<\rho_{t}|\alpha_{t}\>=0$. In other words, $|\alpha_{t}\>$ and $|\rho_{t}\>$
can be considered as part of some orthonormal basis spanning the Hilbert
space. In the spirit of random-matrix theory one can {\em assume}
that the coefficients of an eigenstate $|n\>$ in this basis, and in particular
$\alpha_{t,n}\equiv\<n|\alpha_{t}\>$ and $\rho_{t,n}\equiv\<n|\rho_{t}\>$ are
uncorrelated,
\begin{equation}\label{assume}
P(\rho_{t,n},\alpha_{t,n})=P(\rho_{t,n})\,P(\alpha_{t,n})\,.
\end{equation}
Deferring a discussion of the validity of Eq.~(\ref{assume}) we first point
out its implications. We multiply the identity  
\begin{eqnarray}\label{ident}
H_{0,n}
&=&|\<n|\hat U^{t}|\alpha_{0}\>|^2
\\
&=&|a_{t}|^2\,|\alpha_{t,n}|^2+
2\Re\,a^{*}_{t}r_{t}
\alpha_{t,n}^{*}\rho_{t,n}+|r_{t}|^2|\rho_{t,n}|^2
\nonumber
\end{eqnarray}
with $H_{t,n}=|\alpha_{t,n}|^2$ and average over $n$. Then the second term on
the r.h.s. vanishes since, according to Eq.~(\ref{assume}), the phases from
$\alpha_{t,n}^{*}$ and $\rho_{t,n}$ are uncorrelated. Also the third term
factorizes and gives
$\<|\alpha_{t,n}|^2\>_{n}\,|r_{t}|^2\<|\rho_{t,n}|^2\>_{n}=(1-|a_{t}|^2)/N^2$
after using the normalization implied by Eq.~(\ref{itercoh}). Substitution of
these results into Eq.~(\ref{denscorr}) yields after straightforward
calculation
\begin{eqnarray}
C_{H}(\xi_{0};\xi_{t})
&=&|a_{t}|^2\sqrt{(N\I_{t}-1)/(N\I_{0}-1)}
\label{corripr}
\\&\sim&|a_{t}|^2\,,
\label{result}
\end{eqnarray}
where $\I_{t}=\sum_{n}H_{t,n}^2$ is the inverse participation number in Husimi
representation. The approximation $\I_{0}\sim\I_{t}$ leading to from
(\ref{corripr}) to (\ref{result}) is justified since deviations from the
constant $\I=2/N$ expected within RMT are mainly due to the
influence of periodic orbits \cite{Hel84} which is approximately equal if two
points $\xi_{0}$ and $\xi_{t}$ connected by a short trajectory.

Eq.~(\ref{result}) suggests that also $C_{H}(\xi_{0};\xi_{1})$ is given by
Eq.~(\ref{a1}). This is confirmed by the data in Figs.~\ref{fig2}d and \ref{fig3}
(error bars vs solid line) if the conspicuous sharp curves of very large
correlation (Fig.~\ref{fig2}d) are ignored. To understand the origin of such
exceptional points where the above theory fails we return now to the crucial
step in the derivation, Eq.~(\ref{assume}). A RMT assumption like this is
justified only after all important non-generic effects have been accounted
for. These include in particular symmetries and semiclassical contributions
from short trajectories. Hence, as $|\alpha_{t}\>$ and $|\rho_{t}\>$ are both
concentrated near $\xi_{t}$, we expect that Eq.~(\ref{assume}) breaks down
whenever this point is on a symmetry line or on a periodic orbit. For example,
if $\xi_{0}=\xi_{t}$ we have obviously $C_{H}(\xi_{0};\xi_{t})=1$ but
$|a_{t}|^2\ll 1$ if the orbit is sufficiently unstable. In our model the only
symmetry is time reversal: the unitary transformation $\tilde U=D^\dagger U D$
with $D=\exp(-\i V/2\hbar)$ leads to a symmetric propagator $\tilde U=\tilde
U^{T}$. As a consequence the eigenfunctions of $\tilde U$ are real in position
representation and their Husimi density is symmetric with respect to $p=0$.
After back transformation to the original basis we infer from this symmetry
that the Husimi density at $\xi$ is correlated with the density at
\begin{equation}\label{trpoint}
\tilde\xi=(x,-p-V'(x))
\end{equation}
which is confirmed in Fig.~\ref{fig2}b. For the torus map Eq.~(\ref{map}) the
point $\xi_{t}$ is invariant under this transformation if $[2p_{t}+V'(x_{t})]
\mod 1=0$. Indeed we obtain from this condition with $t=1$ the correct functional
equation for the lines of excess correlation in Fig.~\ref{fig2}d,
\begin{eqnarray}\label{line}
[2p_{0}-V'(x_{0}+p_{0})]\mod 1&=& 0.
\end{eqnarray}
Note that short periodic orbits of period
$1$ and $2$ must be invariant under time reversal and are located on
these lines. In other numerical data for maps without time-reversal invariance
we found excess correlations only at isolated points corresponding to the short
orbits. A quantitative and more detailed analysis of excess correlations is
still to be done. We expect that this will be feasible by applying Gaussian
wavepacket dynamics again, now in order to correlate the states $|\alpha_{t}\>$
and $|\rho_{t}\>$ semiclassically.

We end this paper with a remark on the parameter $|a_{t}|^2$ which essentially
determines the Husimi correlations. Fig.~\ref{fig3} shows that it can be very
large even in a completely chaotic system. In fact $|a_{1}|^2$
(Eq.~(\ref{a1})) is not directly related to the local stability eigenvalues of
the classical map. As explained after Eq.~(\ref{transdp}), this changes only
when the width of the Gaussian is much larger than that of a coherent state, i.e.,
when $|a_{t}|^2$ is already very small.

\begin{acknowledgments} I would like to thank T.~Dittrich, L.~Hufnagel and L.~Kaplan
for discussions.
\end{acknowledgments}

\begin{thebibliography}{33}
\bibitem[{fn1()}]{fn1}
\bibinfo{note}{This is true for systems with orthogonal and unitary symmetry
  \protect\cite{Ber77,Pri95,HS98} but see \protect\cite{FE96} for the
  transition.}
\bibitem[{fn3()}]{fn3}
\bibinfo{note}{This eliminates complications in the theory which are due to the
  compact phase space of torus maps (discrete value of position and momentum,
  periodicity of the wave function in $x$, $p$).}
\end{thebibliography}

\end{document}